\documentclass{ifacconf}

\usepackage{graphicx}      
  \DeclareGraphicsExtensions{.jpeg,.png,.jpg,.eps}
\usepackage{morefloats}
\usepackage{natbib}        
\usepackage{amsmath,amssymb,amsfonts}
\usepackage{color}

\def\eb{{\bf e}}

\def\v{{\bf v}}

\def\S{{\cal S}}

\def\R{{\mathbb R}}

\def\D{\Delta}

\def\l{\lambda}

\def\ap{\rightarrow}

\def\nm{\Vert}

\newcommand{\and}{\mbox{$\wedge$}}

\newcommand{\bc}{\begin{center}}
\newcommand{\ec}{\end{center}}
\newcommand{\be}{\begin{equation}}
\newcommand{\ee}{\end{equation}}
\newcommand{\bd}{\begin{displaymath}}
\newcommand{\ed}{\end{displaymath}}
\newcommand{\ba}{\begin{array}}
\newcommand{\ea}{\end{array}}
\newcommand{\ben}{\begin{enumerate}}
\newcommand{\een}{\end{enumerate}}
\newcommand{\bit}{\begin{itemize}}
\newcommand{\eit}{\end{itemize}}
\newcommand{\beq}{\begin{eqnarray}}
\newcommand{\eeq}{\end{eqnarray}}
\newcommand{\btab}{\begin{tabular}}
\newcommand{\etab}{\end{tabular}}
\newcommand{\bfig}{\begin{figure}}
\newcommand{\efig}{\end{figure}}
\newcommand{\btp}{\begin{tikzpicture}}
\newcommand{\etp}{\end{tikzpicture}}

\def\nmsl1{\nm_{{\rm SL1}}}

\begin{document}
\begin{frontmatter}

\title{Continuous Hands-off Control by \\
CLOT Norm Minimization
} 


\author[NC]{Niharika Challapalli}
\author[MN]{Masaaki Nagahara}
\author[MV]{Mathukumalli Vidyasagar}

\address[NC]{
Department of Electrical Engineering
University of Texas at Dallas
Richardson, TX 75080, USA
(e-mail: niharika15c@gmail.com)}
\address[MN]{
Institute of Environmental Science and Technology,
The University of Kitakyushu,
Hibikino 1-1, Wakamatsu-ku, Kitakyushu, Fukuoka
808-0135, JAPAN (e-mail: nagahara@ieee.org).}
\address[MV]{Department of Systems Engineering,
University of Texas at Dallas,
Richardson, TX 75080, USA, and
Department of Electrical Engineering,
Indian Institute of Technology Hyderabad,
Kandi, Telangana, India 502285 
(e-mail: m.vidyasagar@utdallas.edu, m.vidyasagar@iith.ac.in)}

\begin{abstract}                
In this paper, we consider hands-off control via minimization of the CLOT
(Combined $L$-One and Two) norm.
The maximum hands-off control is the $L^0$-optimal (or the sparsest) control among all feasible controls
that are bounded by a specified value and transfer the state from a given initial state to the origin within a fixed time duration.
In general, the maximum hands-off control is a bang-off-bang control taking values of $\pm 1$ and $0$.
For many real applications, such discontinuity in the control is not desirable.
To obtain a continuous but still relatively
sparse control, we propose to use the CLOT norm,
a convex combination of $L^1$ and $L^2$ norms.
We show by numerical simulation that the CLOT control is continuous and much sparser
(i.e. has longer time duration on which the control takes 0) than the conventional EN (elastic net) control,
which is a convex combination of $L^1$ and squared $L^2$ norms.
\end{abstract}


\end{frontmatter}

\section{Introduction}
\label{sec:Introduction}

Sparsity has recently emerged as an important topic in signal/image processing, machine learning, statistics, etc.
If $y \in \R^m$ and $A \in \R^{m \times n}$ are specified with $m < n$, then
the equation $y = Ax$ is underdetermined and has infinitely many solutions
for $x$ if $A$ has rank $m$.
Finding the sparsest solution (that is, the solution with the fewest
number of nonzero elements) can be formulated as
\[
 \min_{z} \|z\|_0 \mathrm{~~subject~to~~} Az=b.
\]
However, this problem is NP hard, as shown in \citep{Nat95}.
Therefore other approaches have been proposed for this purpose.
This area of research is known as ``sparse regression.''
One of the most popular is LASSO \citep{Tib96}, also referred to as
basis pursuit \citep{CheDonSau99}, in which the $\ell^0$-norm is
replaced by the $\ell^1$-norm.
Thus the problem becomes
\[
 \min_{z} \|z\|_1 \mathrm{~~subject~to~~} Az=b.
\]
The advantage of LASSO is that it is a convex optimization problem and
therefore very large problems can be solved efficiently, for example
by using the \texttt{Matlab}-based package \texttt{cvx} \citep{GraBoy14}.
Moreover, under mild technical assumptions, the LASSO-optimal solution
has no more than $m$ nonzero components \citep{Osborne-Presnell-Turlach00}.
However, the exact location of the nonzero components is very sensitive
to the vector $y$.
To overcome this deficiency, another approach known as the Elastic Net
was proposed in \citep{ZouHas05}, where the $\ell^1$ norm in LASSO
is replaced by
a weighted sum of $\ell^1$ and squared $\ell^2$ norms.
This leads to the optimization problem
\[
 \min_{z} \lambda_1\|z\|_1+\lambda_2\|z\|_2^2 \mathrm{~~subject~to~~} Az=b,
\]
where $\lambda_1$ and $\lambda_2$ are positive weights such that $\lambda_1+\lambda_2=1$.
It is shown in \citep[Theorem 1]{ZouHas05}
that the EN formulation gives the \emph{grouping effect};
If two columns of the matrix $A$ are highly correlated, then the
corresponding components of the solution for $x$ have nearly equal values.
This ensures that the solution for $x$ is not overly sensitive to small
changes in $y$.
The name ``elastic net'' is meant to suggest
a stretchable fishing net that retains \emph{all the big fish}.

During the past decade and a half, another research area known as
``compressed sensing'' has witnessed a great deal of interest.
In compressed sensing, the matrix $A$ is not specified; rather, the user
gets to choose the integer $m$ (known as the number of measurements),
as well as the matrix $A$.
The objective is to choose the matrix $A$ as well as a corresponding
``decooder'' map $\D: \R^m \ap \R^n$ such that, 
the unknown vector $x$ is sparse and the measurement vector $y$
equals $Ax$, then $\D(Ax) = x$ for all sufficiently sparse vectors $x$.
More generally,
if measurement vector $y = Ax + \eta$ where $\eta$ is the
measurement noise, and the vector $x$ is nearly sparse (but not exactly
sparse),
then the recovered vector $\D(Ax + \eta)$ should be sufficiently
close to the true but unknown vector $x$.
This is referred to as ``robust sparse recovery.''
Minimizing the $\ell_1$-norm is among the more popular decoders.
See the books by \citep{Ela}, \citep{EldKut}, and \citep{FouRau} for the
theory and some applications.
Due to its similarity to the LASSO formulation of \citep{Tib96},
this approach to compressed sensing is also referred to as LASSO.

Until recently the situation was that LASSO achieves robust sparse recovery
in compressed sensing, but did not achieve the grouping effect in
sparse regression.
On the flip side, EN achieves the grouping effect, but it was not known
whether it achieves robust sparse recovery.
A recent paper \citep{AhsChaVid16} sheds some light on this problem.
It is shown in \citep{AhsChaVid16} that EN \textit{does not achieve}
robust sparse recovery.
To achieve both the grouping effect in sparse regression as well as
robust sparse recovery in compressed sensing,
\citep{AhsChaVid16} has proposed the CLOT (Combined $L$-One and Two) formulation:
\[
 \min_{z} \lambda_1\|z\|_1+\lambda_2\|z\|_2 \mathrm{~~subject~to~~} Az=b
\mbox{ and } \l_1 + \l_2 = 1 .
\]
The difference between EN and CLOT is the $\ell^2$ norm term;
EN has the squared $\ell^2$ norm while CLOT has the pure $\ell^2$ norm.
This slight change leads to both the grouping effect and robust sparse recovery,
as shown in \citep{AhsChaVid16}.

In parallel with these advances in sparse regression and recovery of
unknown sparse vectors, 
sparsity techniques have also been applied to control.
Sparsity-promoting optimization has been applied to networked
control in \citep{NagQueOst14}, where
quantization errors and data rate can be reduced at the same time
by sparse representation of control packets.
Other examples of control applications include
optimal controller placement by \citep{CasClaKun12,ClaKun12,FarLinJov11},
design of feedback gains by \citep{FuFarJov13,PolKhlShc13},
state estimation by \citep{ChaAsiRomRoz11},
to name a few. 

More recently, 
a novel control called the {\em maximum hands-off control} has been 
proposed in~\citep{NagQueNes16} for \emph{continuous-time} systems.
The maximum hands-off control 
is the $L^0$-optimal control (the control that has the minimum support length) among all feasible controls
that are bounded by a fixed value and transfer the state from a given initial state to the origin within a fixed time duration.
Such a control is effective for
reduction of electricity or fuel consumption; an electric/hybrid vehicle
shuts off the internal combustion engine (i.e. hands-off control)
when the vehicle is stopped or the speed is lower than a preset threshold; see \citep{Cha07} for example.
Railway vehicles also utilize hands-off control, often called {\em coasting control},
to cut electricity consumption; see \citep{LiuGol03} for details.
In~\citep{NagQueNes16}, the authors have proved the theoretical relation between 
the maximum hands-off control and the $L^1$ optimal control 
under the assumption of normality.
Also, important properties of the maximum hands-off
control have been proved
in \citep{IkeNag16_automatica} for
the convexity of the value function,
and in \citep{ChaNagQueRao16} 
for the existence and the discreteness.

In general, the maximum hands-off control is a bang-off-bang control taking values of $\pm 1$ and $0$.
For many real applications, such a discontinuity property is not desirable.
To obtain a continuous but still sparse control, 
\citep{NagQueNes16} has proposed to use a combined $L^1$ and \emph{squared} $L^2$ minimization,
like EN mentioned above.
Let us call this control an EN control.
As in the case of EN in the vector optimization, the EN control often shows much less sparse
(i.e. has a larger $L^0$ norm) than the maximum hands-off control.
Then, we proposed to use the CLOT norm,
a convex combination of $L^1$ and \emph{non-squared} $L^2$ norms.
We call the minimum CLOT-norm control the CLOT control.
We show by numerical simulation that the CLOT control is continuous and much sparser
(i.e. has longer time duration on which the control takes 0) than the conventional EN control.

The remainder of this article is organized as follows.
In Section \ref{sec:Problem Formulation}, we formulate the control problem considered in this paper.
In Section \ref{sec:Discretization}, we give a discretization method to numerically compute the optimal control.
Results of the numerical computations on a variety of problems
are presented in Section \ref{sec:exam}.
These examples illustrate the advantages of the CLOT control
compared with the maximum hands-off control and the EN control.
We present some conclusions in Section \ref{sec:Conclusions}.

\subsection*{Notation}

Let $T>0$ and $m\in\mathbb{N}$. 
For a continuous-time signal $u(t)\in\mathbb{R}$ over a time interval $[0, T]$, 
we define its $L^p$ ($p\geq 1$) and $L^{\infty}$ norms respectively by
\[
 \|u\|_{p} \triangleq \bigg\{\int_{0}^{T} |u(t)|^{p} dt\bigg\}^{1/p},~
 \|u\|_{\infty} \triangleq \sup_{t\in[0, T]}|u(t)|.
\]
We denote the set of all signals with $\|u\|_{p}<\infty$ by $L^p[0,\,T]$
for $p\geq 1$ or $p=\infty$.
We define the $L^0$ norm of a signal $u(t)$ on the interval $[0, T]$ as
\[
  \|u\|_{0}\triangleq \int_{0}^{T}\phi_{0}(u(t)) dt,
\]
where $\phi_{0}$ is the $L^0$ kernel function defined by
\begin{equation}
  \phi_0(\alpha) \triangleq 
    \begin{cases} 
    1, & \text{~if~} \alpha\neq 0,\\ 
    0, & \text{~if~} \alpha=0
    \end{cases}
\label{eq:L0-kernel}
\end{equation}
for a scalar $\alpha\in\mathbb{R}$.
The $L^0$ norm can be represented by
\[
 \|u\|_0 = \mu_{\mathrm{L}}\bigl(\mathrm{supp}(u)\bigr),
\]
where $\mathrm{supp}(u)$ is the support of the signal $u$, and
$\mu_{\mathrm{L}}$ is the Lebesgue measure on ${\mathbb{R}}$.

\section{Problem Formulation}
\label{sec:Problem Formulation}

Let us consider a linear time-invariant system described by
\begin{equation}
 \frac{dx}{dt}(t) = Ax(t) + Bu(t),~ t\geq 0,~ x(0)=\xi.
 \label{eq:plant}
\end{equation}
Here we assume that $x(t)\in{\mathbb{R}}^n$, $u(t)\in{\mathbb{R}}$,
and the initial state $x(0)=\xi$ is fixed and given.
The control objective is to drive the state $x(t)$ from $x(0)=\xi$ to the origin
at time $T>0$, that is
\begin{equation}
 x(T)=0.
 \label{eq:terminal condition}
\end{equation}
We limit the control $u(t)$ to satisfy
\begin{equation}
 \|u\|_\infty \leq U_{\max}
 \label{eq:u_max}
\end{equation}
for fixed $U_{\max}>0$.

If the system \eqref{eq:plant} is controllable and the final time $T$ is larger than
the optimal time $T^\ast$ (the minimal time in which there exist a control $u(t)$
that drives $x(t)$ from $x(0)=\xi$ to the origin; see \citep{HerLas}), 
then there exists at least one
$u(t)\in L^\infty[0,T]$ that satisfies
equations \eqref{eq:plant}, \eqref{eq:terminal condition}, and \eqref{eq:u_max}.
Let us call such a control a \emph{feasible} control.
From \eqref{eq:plant} and \eqref{eq:terminal condition},
any feasible control $u(t)$ on $[0,T]$ satisfies
\[
 0=x(T)=e^{AT}\xi + \int_0^T e^{A(T-t)}Bu(t)dt,
\]
or
\begin{equation}
 \int_0^T e^{-At}Bu(t)dt+\xi=0.
 \label{eq:feasible control}
\end{equation}
Define a linear operator $\Phi:L^\infty[0,T] \rightarrow {\mathbb{R}}^n$ by
\[
 \Phi u \triangleq \int_0^T e^{-At}Bu(t)dt,\quad u\in L^\infty[0,T].
\]
By this, we define the set ${\mathcal{U}}$ of the feasible controls by
\begin{equation}
 \mathcal{U}\triangleq
  \left\{u\in L^\infty: \Phi u+ \xi=0,~\|u\|_\infty\leq 1\right\}.
  \label{eq:feasible controls}
\end{equation}

The problem of the maximum hands-off control
is then described by
\begin{equation}
 \min_{u} \|u\|_0 \mathrm{~~subject~to~~} u \in \mathcal{U}.
 \label{eq:L0 optimal control}
\end{equation}

The $L^0$ problem \eqref{eq:L0 optimal control} is very hard to solve since the $L^0$ cost function
is non-convex and discontinuous.
For this problem, \citep{NagQueNes16} has shown that the $L^0$ optimal control in \eqref{eq:L0 optimal control} is
equivalent to the following $L^1$ optimal control:
\begin{equation}
 \min_{u} \|u\|_1 \mathrm{~~subject~to~~} u \in \mathcal{U},
 \label{eq:LASSO}
\end{equation}
\emph{if} the plant is normal, that is, if the \eqref{eq:plant} is controllable and the matrix $A$ is nonsingular.
Let us call the $L^1$ optimal control as the \emph{LASSO control}.
If the plant is normal, then the LASSO control is in general a \emph{bang-off-bang} control
that is piecewise constant taking values in $\{0,\pm 1\}$.
The discontinuity of the LASSO solution is not desirable in real applications,
and a smoothed solution is also proposed in \citep{NagQueNes16} as
\begin{equation}
 \min_{u} \|u\|_1+\lambda\|u\|_2^2 \mathrm{~~subject~to~~} u \in \mathcal{U},
 \label{eq:EN}
\end{equation}
where $\lambda>0$ is a design parameter for smoothness.
Let us call this control the \emph{EN (elastic net) control}.
In \citep{NagQueNes16}, it is proved that the solution of \eqref{eq:EN} is
a continuous function on $[0,T]$.

While the EN control is continuous, it is shown by numerical experiments that
the EN control is not sometimes sparse.
This is an analogy of the EN for finite-dimensional vectors that
EN does not achieve robust sparse recovery.
Borrowing the idea of CLOT in \citep{AhsChaVid16},
we define the CLOT optimal control problem by
\begin{equation}
 \min_{u} \|u\|_1+\lambda\|u\|_2 \mathrm{~~subject~to~~} u \in \mathcal{U}.
 \label{eq:CLOT}
\end{equation}
We call this optimal control the \emph{CLOT control}.

\section{Discretization}
\label{sec:Discretization}

Since the problems \eqref{eq:LASSO}--\eqref{eq:CLOT} are infinite dimensional,
we should approximate it to finite dimensional problems.
For this, we adopt the time discretization.

First, we divide the time interval $[0,T]$ into $N$ subintervals,
$[0,T] = [0,h) \cup \dots \cup [(N-1)h,Nh]$,
where $h$ is the discretization step (or the sampling period)
such that $T=Nh$.
We assume that the state $x(t)$ and the control $u(t)$ in \eqref{eq:plant}
are constant over each subinterval.
On the discretization grid,
$t=0,h,\dots,Nh$,
the continuous-time system \eqref{eq:plant} is described as
\begin{equation}
 x_d[m+1] = A_d x_d[m] + B_d u_d[m],\quad m=0,1,\dots,N-1,
\end{equation}
where $x_d[m]\triangleq x(mh)$, $u_d[m]\triangleq u(mh)$, and
\begin{equation}
 A_d \triangleq e^{Ah},\quad B_d \triangleq \int_0^h e^{At}Bdt.
\end{equation}
Define the control vector
\begin{equation}
 \vec{u}_d \triangleq [u_d[0], u_d[1],\dots,u_d[N-1]]^\top.
\end{equation} 
Note that the final state $x(T)$ can be described as
\begin{equation}
 x(T)=x_d[N]=A_d^N \xi + \Phi_N \vec{u}_d,
\end{equation}
where
\begin{equation}
 \Phi_N \triangleq \begin{bmatrix}A_d^{N-1}B_d,&A_d^{N-2}B_d,&\dots,&B_d\end{bmatrix}.
 \label{eq:Phi_N}
\end{equation}
Then the set $\mathcal{U}$ in \eqref{eq:feasible controls} is approximately
represented by
\begin{equation}
 {\mathcal{U}}_N \triangleq
 \left\{\vec{u}_d\in {\mathbb{R}}^N: A_d^N\xi + \Phi_N \vec{u}_d =0,~\|\vec{u}_d\|_\infty\leq 1\right\}.
 \label{eq:approximated feasible controls}
\end{equation}

Next, we approximate the $L^1$ norm of $u$ by
\begin{equation}
 \begin{split}
 \|u\|_1 &= \int_0^T |u(t)| dt\\
 &= \sum_{m=0}^{N-1} \int_{mh}^{(m+1)h} |u(t)| dt\\
 &\approx \sum_{m=0}^{N-1} \int_{mh}^{(m+1)h} |u_d[m]| dt\\
 &= \sum_{m=0}^{N-1} |u_d[m]| h\\
 &= \|\vec{u}_d\|_1 h.
 \end{split}
\end{equation}
In the same way, we obtain approximation of the $L^2$ norm of $u$ as
\begin{equation}
\|u\|_2^2 = \int_0^T |u(t)|^2 dt\approx \|\vec{u}_d\|_2^2 h.
\end{equation}

Finally, the optimal control problems \eqref{eq:LASSO},
\eqref{eq:EN} and \eqref{eq:CLOT} can be approximated by
\begin{equation}
 \min_{\vec{u}_d\in{\mathbb{R}}^N} h\|u\|_1 \mathrm{~~subject~to~~} \vec{u}_d\in{\mathcal{U}}_N
 \label{eq:LASSO_d}
\end{equation}
\begin{equation}
 \min_{\vec{u}_d\in{\mathbb{R}}^N} h\|\vec{u}_d\|_1+h\lambda\|\vec{u}_d\|_2^2 \mathrm{~~subject~to~~} \vec{u}_d\in{\mathcal{U}}_N
 \label{eq:EN_d}
\end{equation}
\begin{equation}
 \min_{\vec{u}_d\in{\mathbb{R}}^N} h\|\vec{u}_d\|_1+\sqrt{h}\lambda \|\vec{u}_d\|_2 \mathrm{~~subject~to~~} \vec{u}_d\in{\mathcal{U}}_N
 \label{eq:CLOT_d}
\end{equation}
The optimization problems
are convex and can be efficiently solved by
numerical software packages such as \verb=cvx= with Matlab;
see \citep{GraBoy14} for details.

\section{Numerical Examples}
\label{sec:exam}

In this section we present numerical results from applying the CLOT norm
minimization approach to seven different plants, and compare the results
with those from applying LASSO and EN.

\subsection{Details of Various Plants Studied}

For the reader's convenience, the details of the various plants are
given in Table \ref{table:plants}.
The figure numbers show where the corresponding
computational results can be found.
Some conventions are adopted to reduce the clutter in the table, as
described next.
All plants are of the form
\bd
P(s) = \frac{n(s)}{d(s)}, n(s) = \prod_{i=1}^{n_z} (s - z_i ) ,
d(s) = \prod_{i=1}^{n_p} (s - p_i) .
\ed
To save space in the table, the plant zeros are not shown;
$P_3(s)$ has a zero at $s = -2$, $P_6(s)$ has a zero at $s = 2$,
while $P_7(s)$ has zeros at $s = 1, 2$.
The remaining plants do not have any zeros, so that the plant
numerator equals one.

Once the plant zeros and poles are specified, the plant numerator
and denominator polynomials $n,d$ were computed using the Matlab command
{\tt poly}.
Then the transfer function was computed as {\tt P = tf(n,d)},
and the state space realization was computed as
{\tt [A,B,C,D] = ssdata(P)}.
The maximum
control amplitude is taken $1$, so that the control must satisfy
$|u(t)| \leq 1$ for $t \in [0,T]$.
To save space, we use the notation $\eb_l$ to denote an $l$-column
vector whose elements all equal one.
Note that in all but one case, the initial condition equals $\eb_n$
where $n$ is the order of the plant.

Note that, with $T = 20$, the problems with plants $P_6(s)$
and $P_7(s)$ are not feasible (meaning that $T$ is smaller than
the minimum time needed to reach the origin); this is why we took $T = 40$.

All optimization problems were solved after discretizing the interval
$[0,T]$ into both 2,000 as well as 4,000 samples, to examine whether the
sampling time affects the sparsity density of the computed optimal control.

\begin{table}
\centering
\btab{|c|c|c|c|c|c|c|}
\hline
\textbf{No.} & \textbf{Plant} &\textbf{Poles} & $T$ & $x(0)$ & $\l$
& \textbf{Figs.} \\
\hline
1 & $P_1(s)$ & (0,0,0,0) & 20 & $\eb_4$ & 1 & 
\ref{fig:P1_state_l_1} , \ref{fig:P1_control_l_1} \\
\hline
2 & $P_1(s)$ & (0,0,0,0) & 20 & $\eb_4$ & 0.1 &
\ref{fig:P1_state_l_01} , \ref{fig:P1_control_l_01} \\
\hline
3 & $P_2(s)$ & $ -0.025 \pm j$ & 20 & $\eb_2$ & 0.1 &
\ref{fig:P2_state_1_1} , \ref{fig:P2_control_1_1} \\
\hline
4 & $P_2(s)$ & $ -0.025 \pm j$ & 20 & $(10,1)^\top$ & 0.1 &
\ref{fig:P2_state_10_1} , \ref{fig:P2_control_10_1} \\
\hline
5 & $P_3(s)$ & $-1 \pm 0.2j , \pm j$ & 20 & $\eb_4$ & 0.1 & 
\ref{fig:P3_state}, \ref{fig:P3_control} \\
\hline
6 & $P_4(s)$ & $-1 \pm 0.2j , $ & 20 & $\eb_4$ & 0.1 & 
\ref{fig:P4_state}, \ref{fig:P4_control} \\
& & $ -0.3 \pm j $ & & & & \\
\hline
7 & $P_5(s)$ & $-5 \pm j , $ & 20 & $\eb_6$ & 0.1 & 
\ref{fig:P5_state}, \ref{fig:P5_control} \\
& & $-0.3 \pm 2j$ , & & & & \\
& & $-1 \pm 2 \sqrt{2} j$ & & & & \\
\hline
8 & $P_6(s)$ & $0,0,0,0,\pm j$ & 40 & $\eb_6$ & 0.1 & 
\ref{fig:P6_state}, \ref{fig:P6_control} \\
\hline
9 & $P_7(s)$ & $0,0,0,0,\pm j$ & 40 & $\eb_6$ & 0.1 & 
\ref{fig:P7_state}, \ref{fig:P7_control} \\
\hline
\etab
\caption{Details of various plants studied.}
\label{table:plants}
\end{table}

\subsection{Plots of Optimal State and Control Trajectories}

The plots of the $\ell^2$-norm (or Euclidean norm)
of the state vector trajectory and the
control signal for all these examples are shown in the next several plots.

\begin{figure}[t]
\centering
\includegraphics[width=80mm]{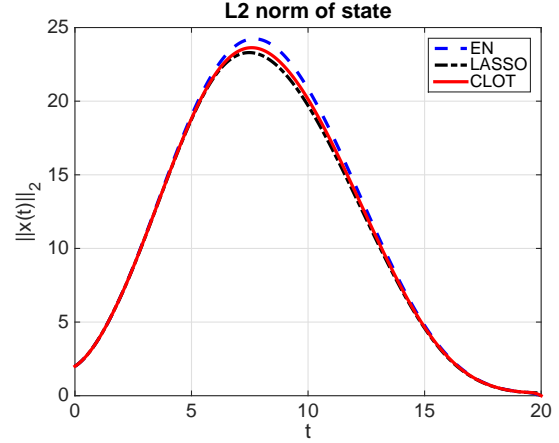}
\caption{State trajectory for the plant $P_1(s)$ with the
initial state $(1,1,1,1)^\top$ and $\l = 1$.}
\label{fig:P1_state_l_1}
\end{figure}

\begin{figure}[t]
\centering
		\includegraphics[width=80mm]{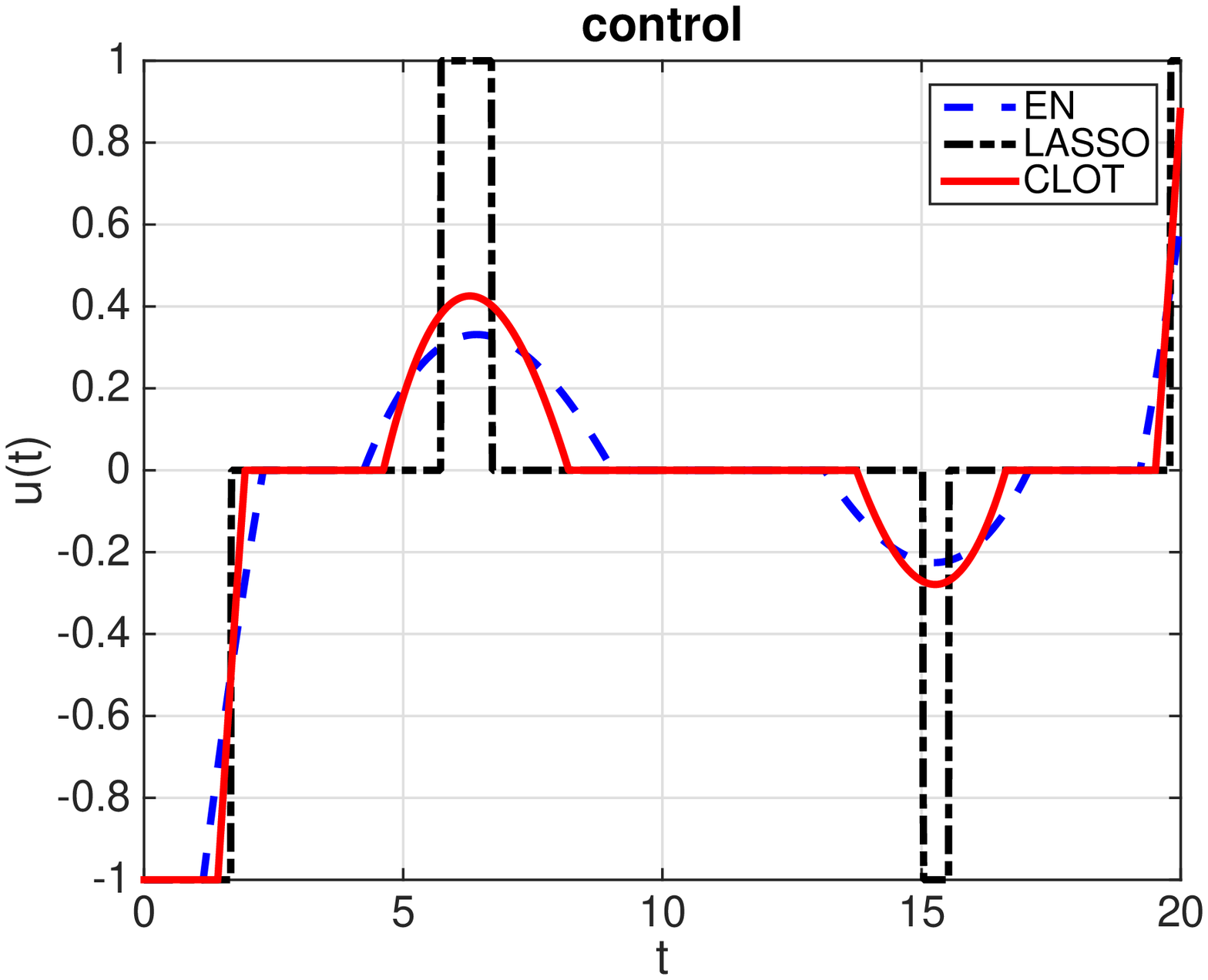}
	\caption{Control trajectory for the plant $P_1(s)$ with the
initial state $(1,1,1,1)^\top$ and $\l = 1$.}
\label{fig:P1_control_l_1}
\end{figure}

\begin{figure}[t]
\centering
		\includegraphics[width=80mm]{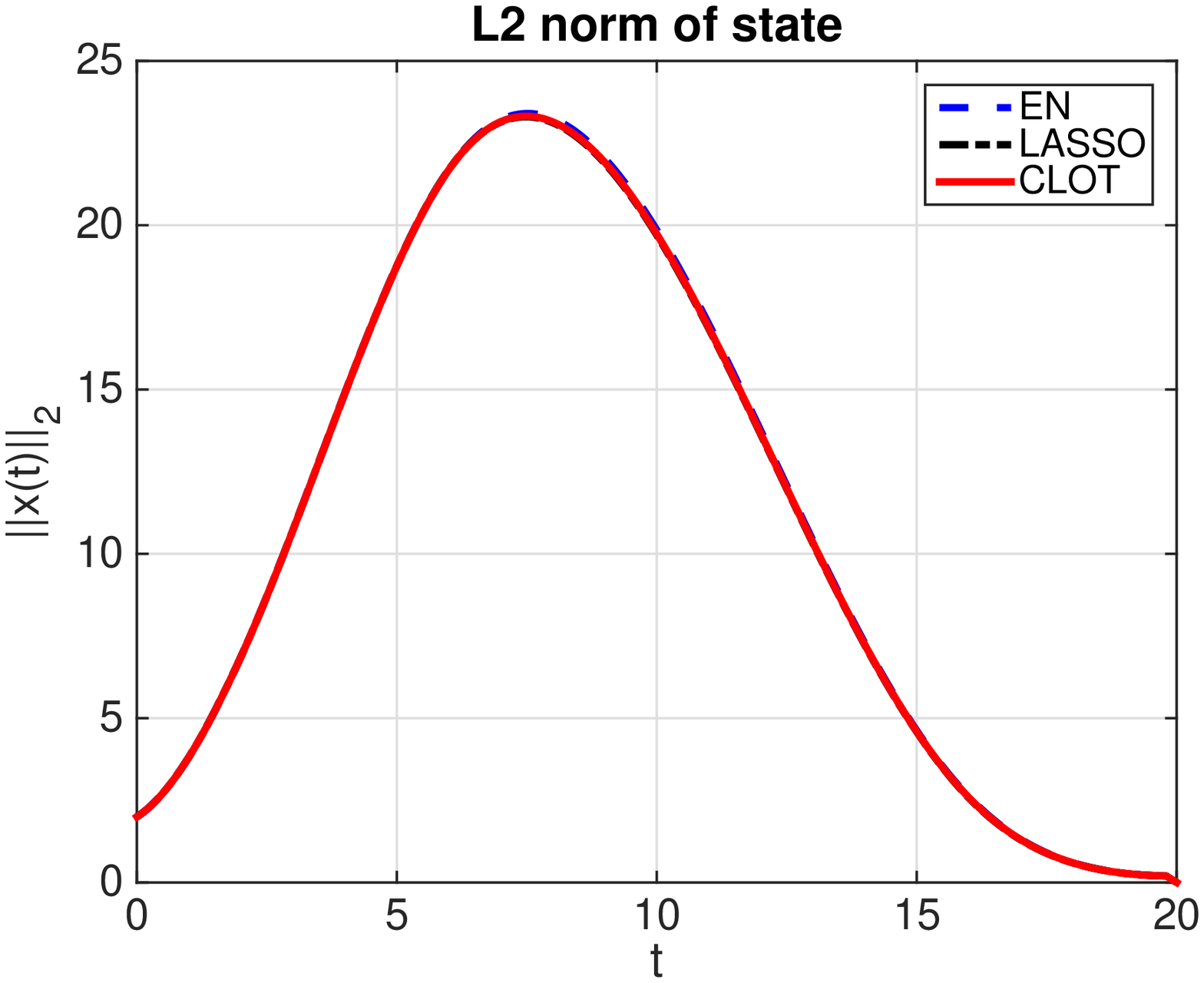}
\caption{State trajectory for the plant $P_1(s)$ with the
initial state $(1,1,1,1)^\top$ and $\l = 0.1$.}
\label{fig:P1_state_l_01}
\end{figure}

\begin{figure}[t]
\centering
		\includegraphics[width=80mm]{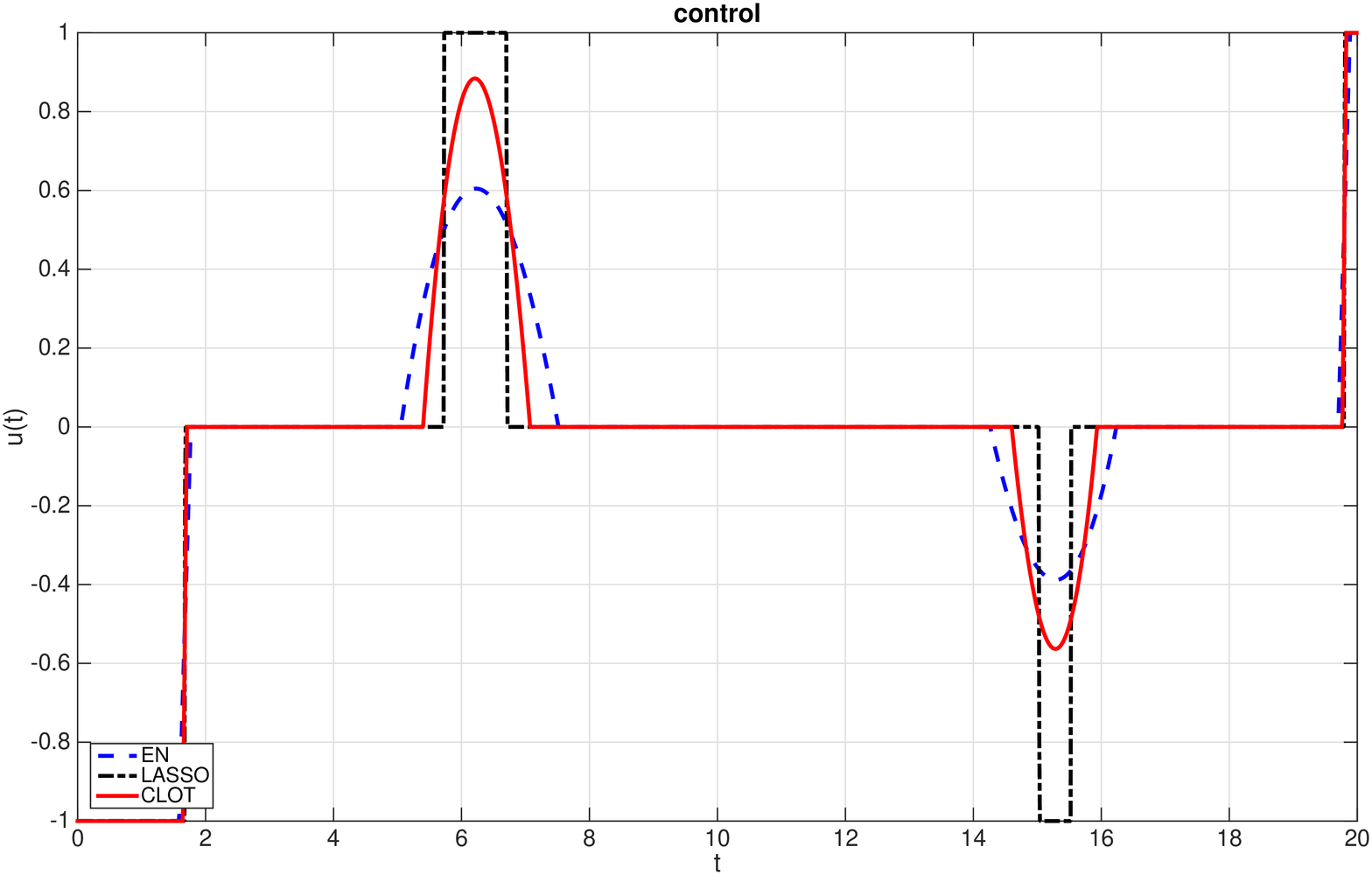}
\caption{Control trajectory for the plant $P_1(s)$ with the
initial state $(1,1,1,1)^\top$ and $\l = 0.1$.}
	\label{fig:P1_control_l_01}
\end{figure}

We begin with the plant $P_1(s)$, the fourth-order integrator.
Figures \ref{fig:P1_state_l_1} and \ref{fig:P1_control_l_1}
show the state and control trajectories when $\l = 1$.
The same system is analyzed using a smaller value of $\l = 0.1$.
One would expect that the resulting control signals would be more sparse
with a smaller $\l$, and this is indeed the case.
The results are shown in 
Figures \ref{fig:P1_state_l_01} and \ref{fig:P1_control_l_01}.
Based on the observation that the control signal becomes more sparse
with $\l = 0.1$ than with $\l = 1$, all the other plants are analyzed
with $\l = 0.1$.
	
\begin{figure}[t]
\centering
		\includegraphics[width=80mm]{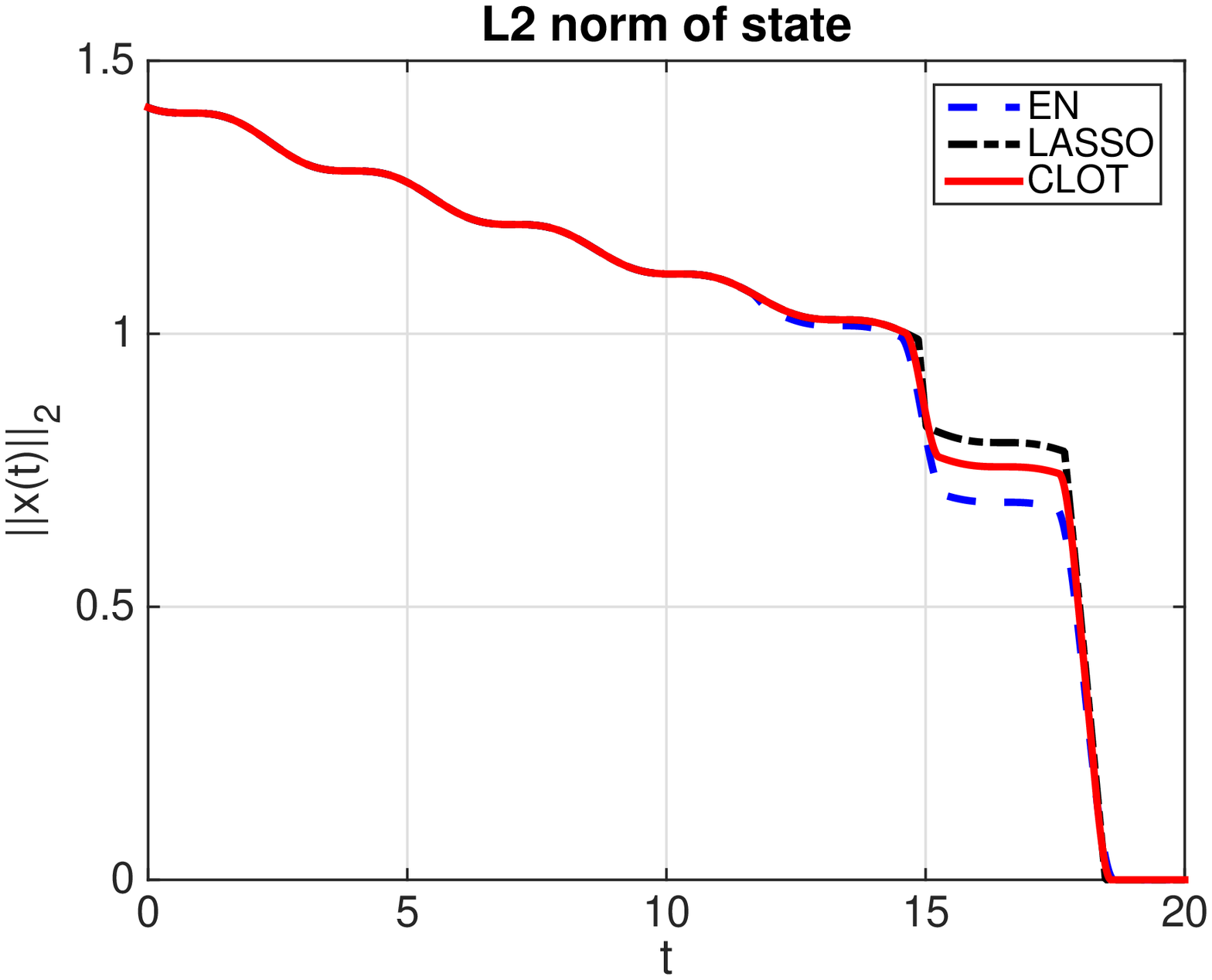}
\caption{State trajectory for the the plant $P_2(s)$ with the
initial state $(1,1)^\top$ and $\l = 0.1$.}
\label{fig:P2_state_1_1}
\end{figure}

\begin{figure}[t]
\centering
		\includegraphics[width=80mm]{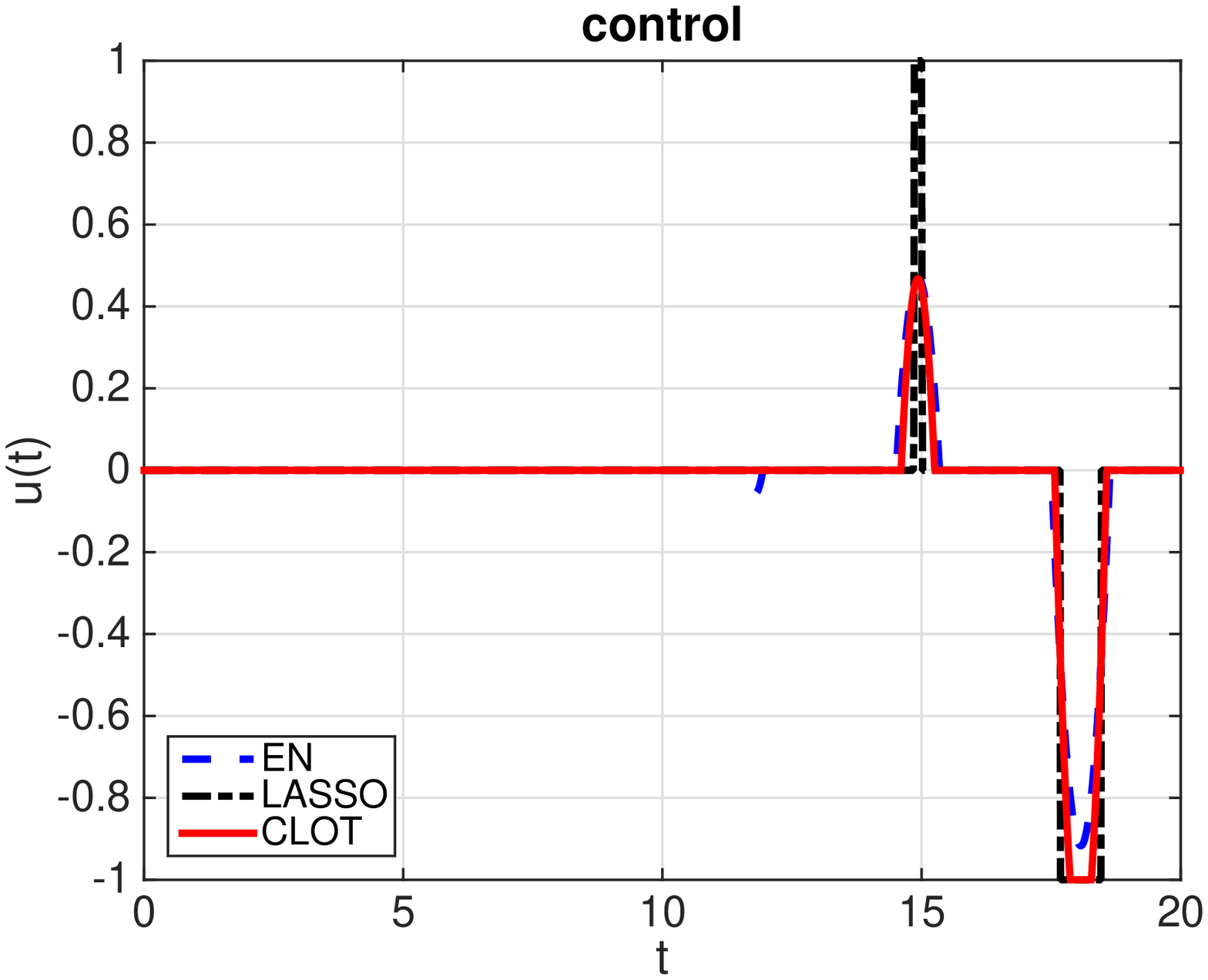}
\caption{Control trajectory for the the plant $P_2(s)$ with the
initial state $(1,1)^\top$ and $\l = 0.1$.}
\label{fig:P2_control_1_1}
\end{figure}

\begin{figure}[t]
\centering
\includegraphics[width=80mm]{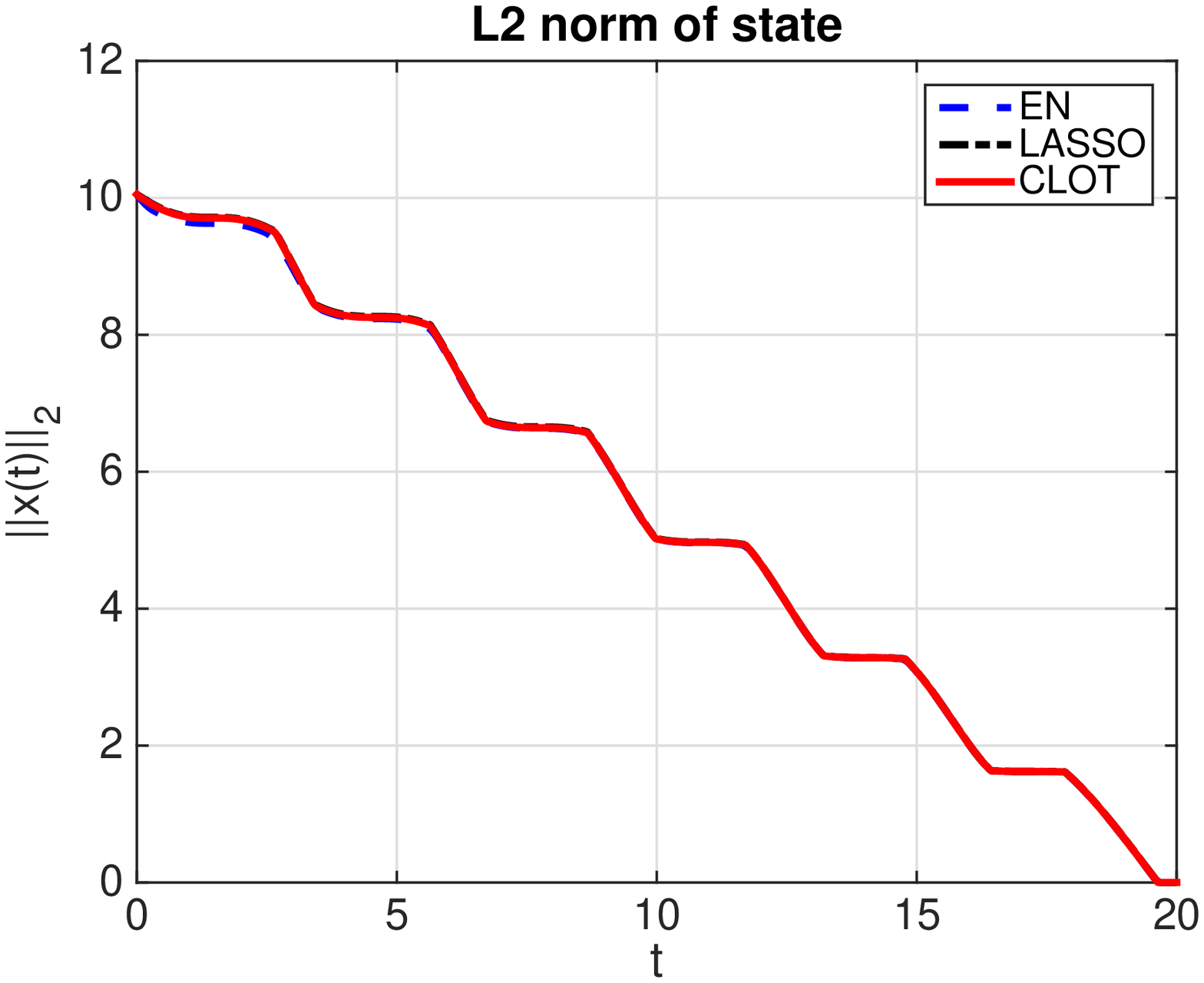}
\caption{State trajectory for the plant $P_2(s)$ with the
initial state $(10,1)^\top$ and $\l = 0.1$.}
\label{fig:P2_state_10_1}
\end{figure}

\begin{figure}[t]
\centering
\includegraphics[width = 80mm]{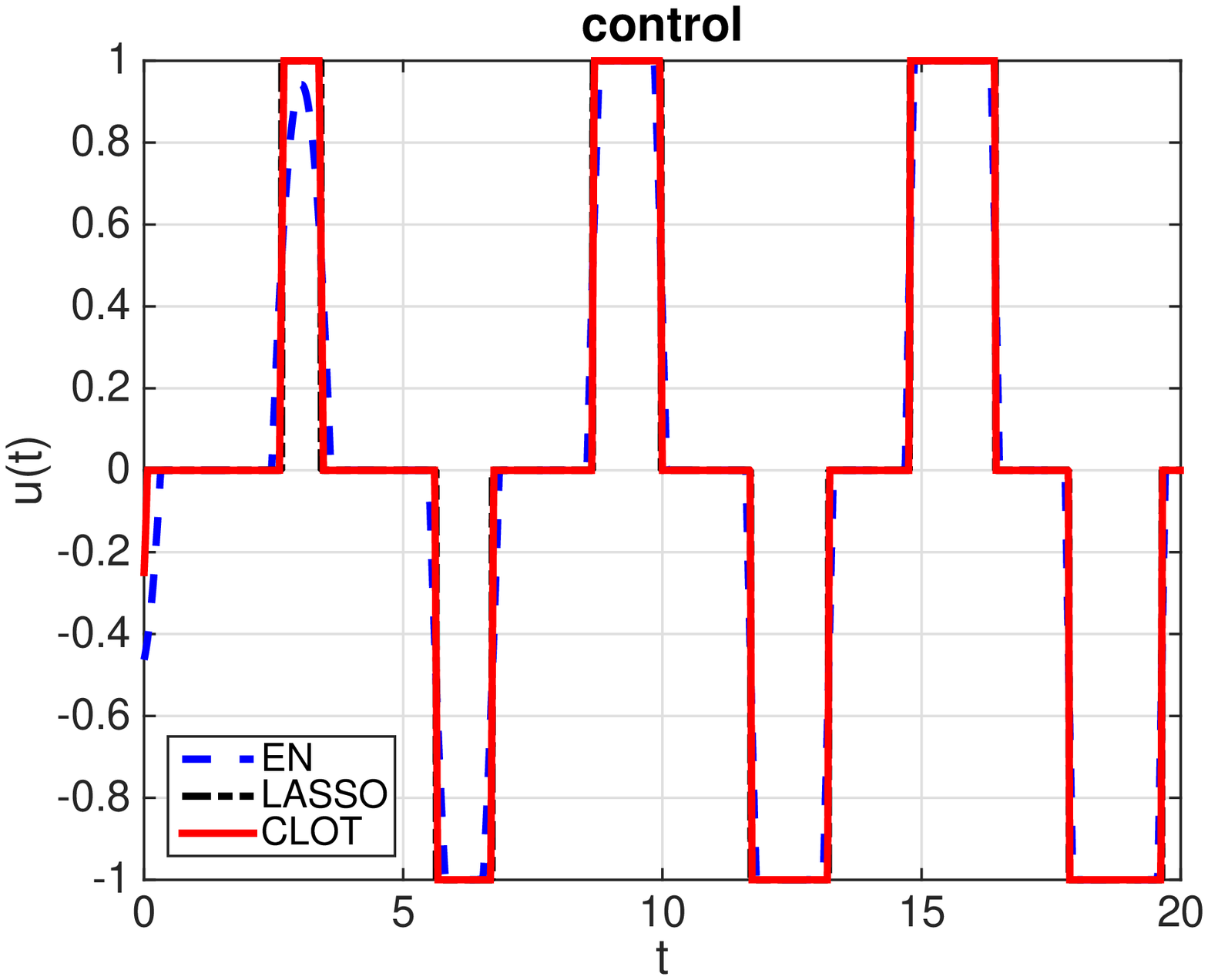}
\caption{Control trajectory for the plant $P_2(s)$ with the
initial state $(10,1)^\top$ and $\l = 0.1$.}
\label{fig:P2_control_10_1}
\end{figure}

Figures \ref{fig:P2_state_1_1} and \ref{fig:P2_control_1_1}
display the state trajectory and the control
trajectories of the plant $P_2(s)$ (damped harmonic oscillator)
when the initial state is $(1,1)^\top$.
Figures \ref{fig:P2_state_10_1} and \ref{fig:P2_control_10_1}
show the state and control trajectories with the initial state
$(10,1)^\top$.
It can be seen that, with this intial state, the control signal
changes sign more frequently.

\begin{figure}[t]
\centering
\includegraphics[width=80mm]{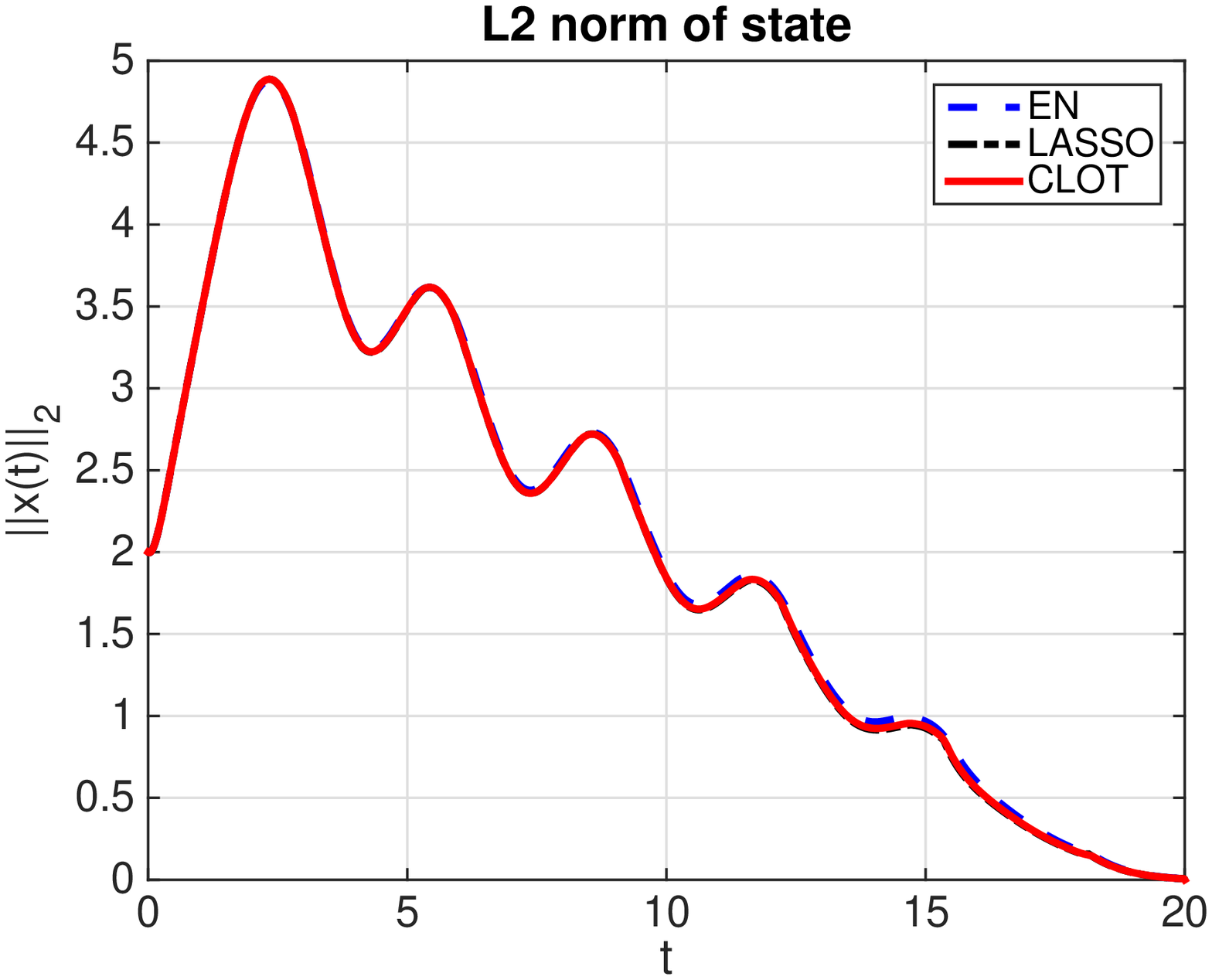}
\caption{State trajectory for the plant $P_3(s)$ with the initial
state $(1,1,1,1)^\top$ and $\l = 0.1$.}
\label{fig:P3_state}
\end{figure}

\begin{figure}[t]
\centering
\includegraphics[width=80mm]{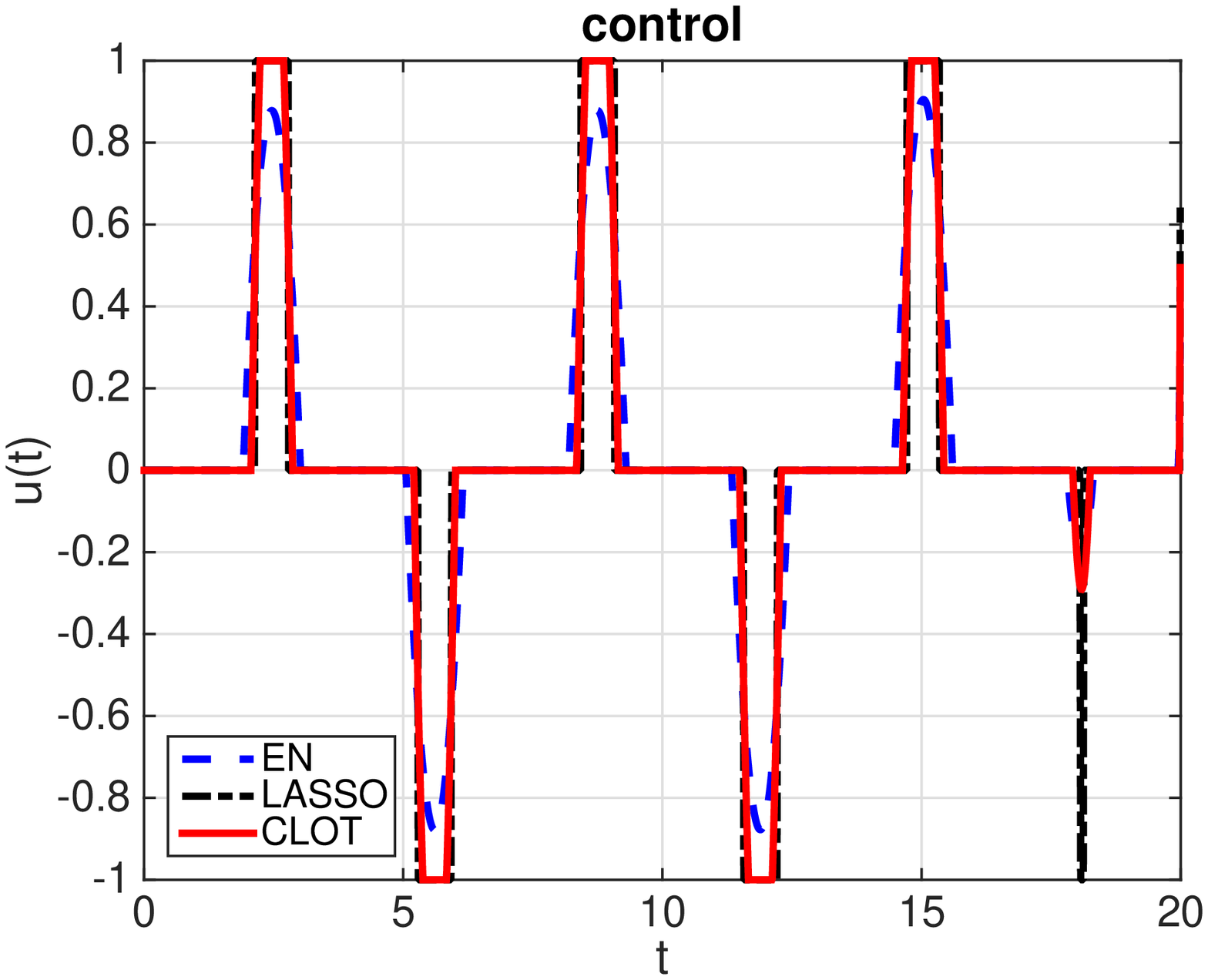}
\caption{Control trajectory for the plant $P_3(s)$ with the initial
state $(1,1,1,1)^\top$ and $\l = 0.1$.}
\label{fig:P3_control}
\end{figure}

\begin{figure}[t]
\centering
\includegraphics[width=80mm]{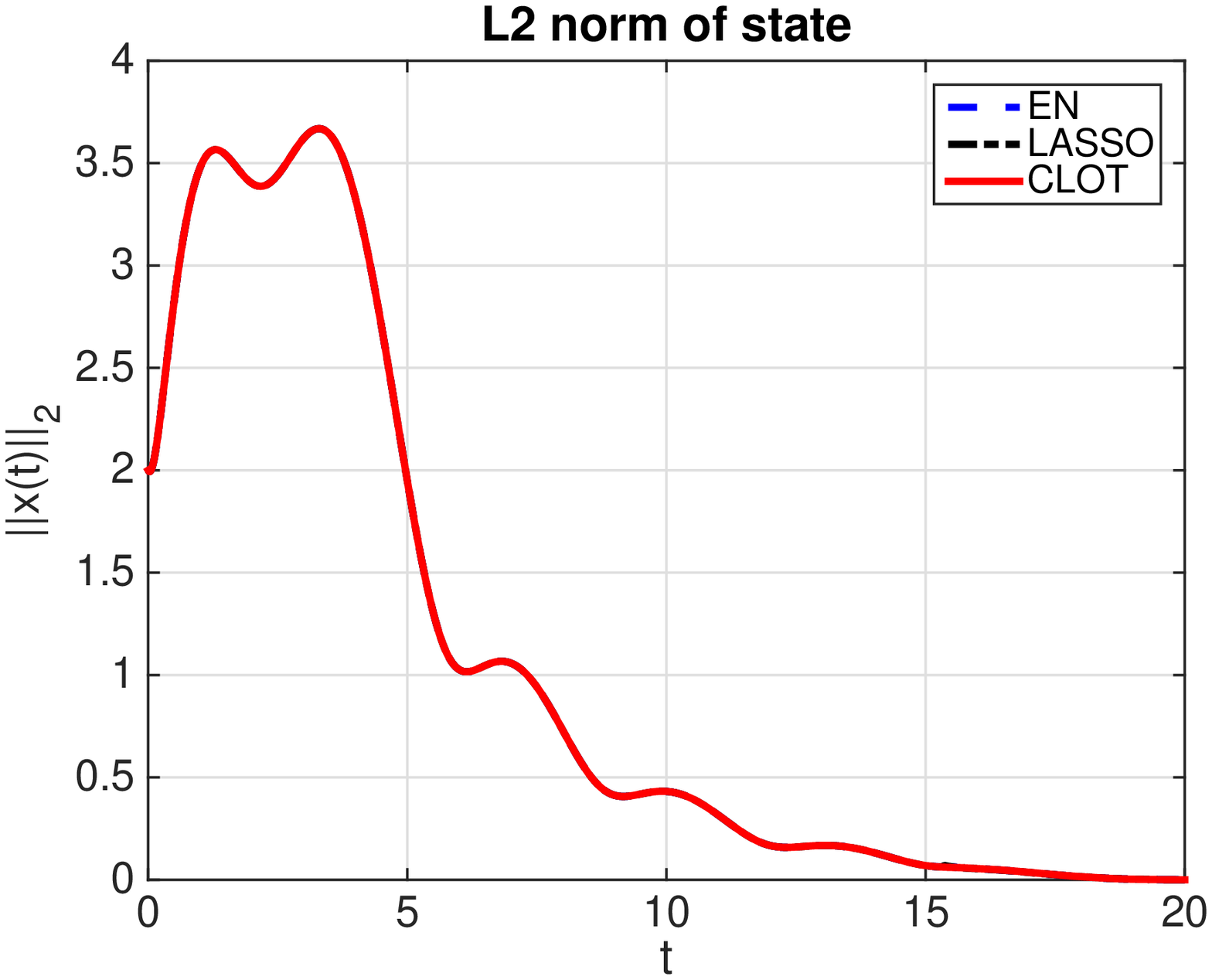}
\caption{State trajectory for the plant $P_4(s)$ with the initial
state $(1,1,1,1)^\top$ and $\l = 0.1$.}
\label{fig:P4_state}
\end{figure}

\begin{figure}[t]
\centering
\includegraphics[width=80mm]{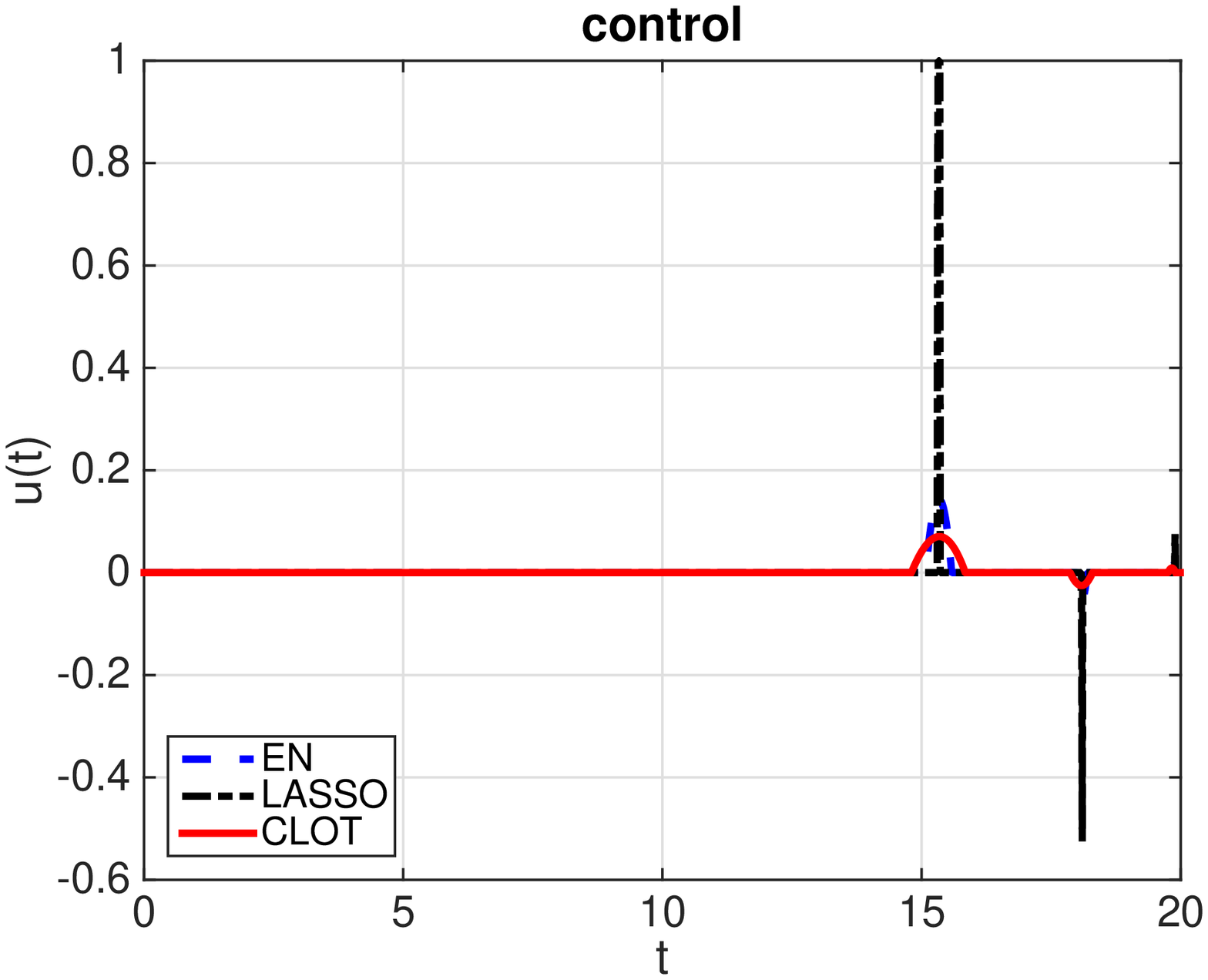}
\caption{Control trajectory for the plant $P_4(s)$ with the initial
state $(1,1,1,1)^\top$ and $\l = 0.1$.}
\label{fig:P4_control}
\end{figure}

\begin{figure}[t]
\centering
\includegraphics[width=80mm]{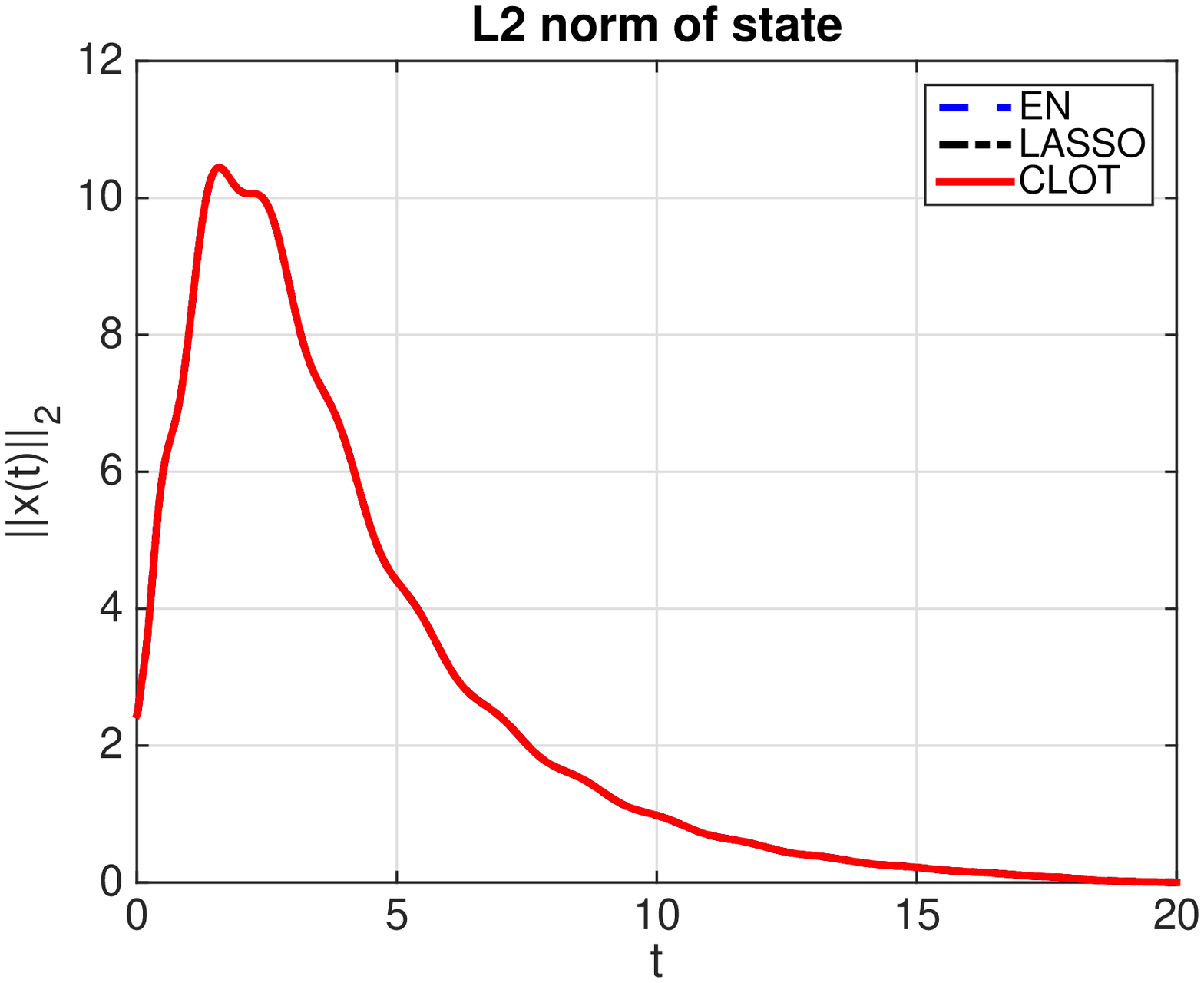}
\caption{State trajectory for the plant $P_5(s)$ with the initial
state $(1,1,1,1,1,1)^\top$ and $\l = 0.1$.}
\label{fig:P5_state}
\end{figure}

\begin{figure}[t]
\centering
\includegraphics[width=80mm]{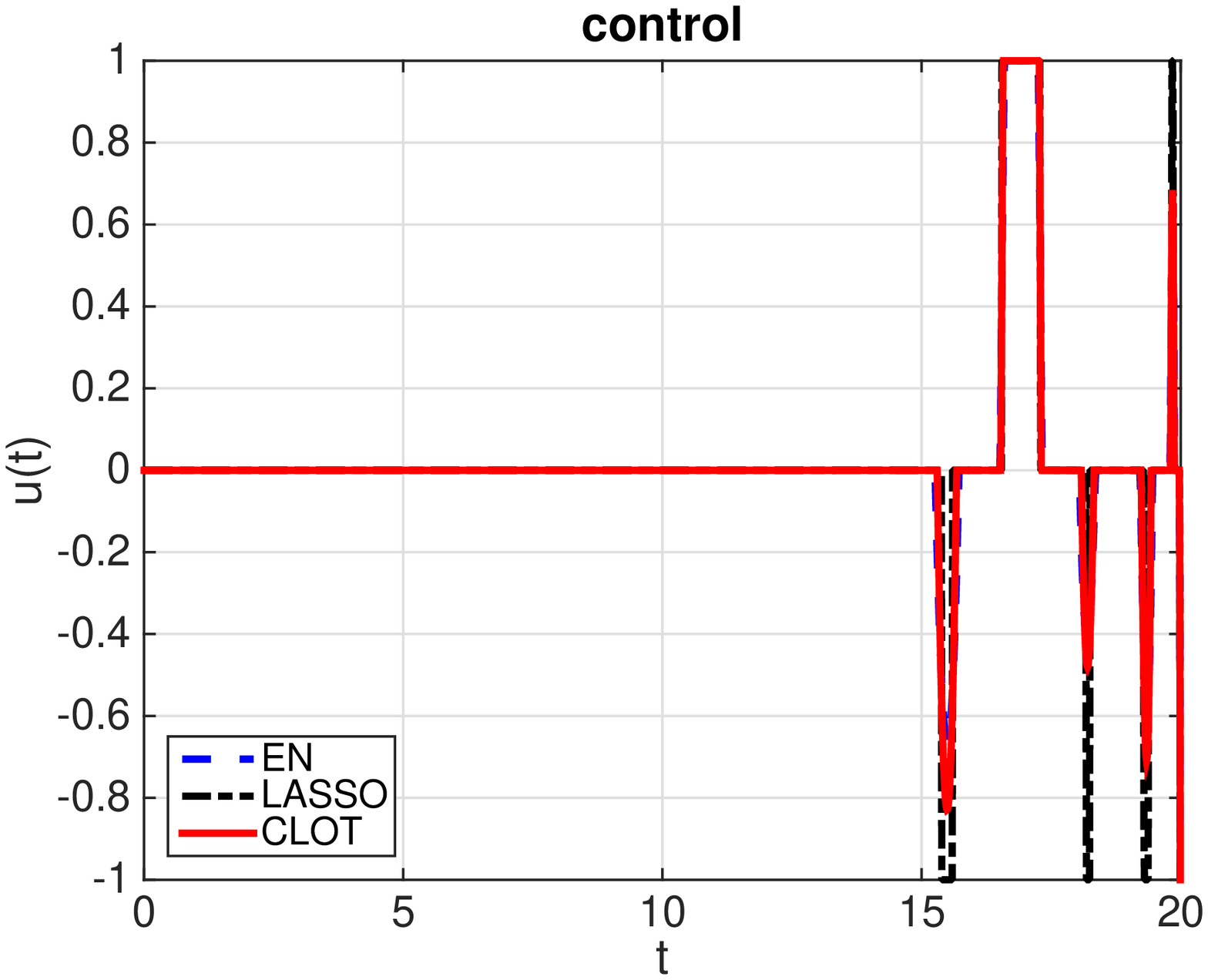}
\caption{Control trajectory for the plant $P_5(s)$ with the initial
state $(1,1,1,1,1,1)^\top$ and $\l = 0.1$.}
\label{fig:P5_control}
\end{figure}

\begin{figure}[t]
\centering
\includegraphics[width=80mm]{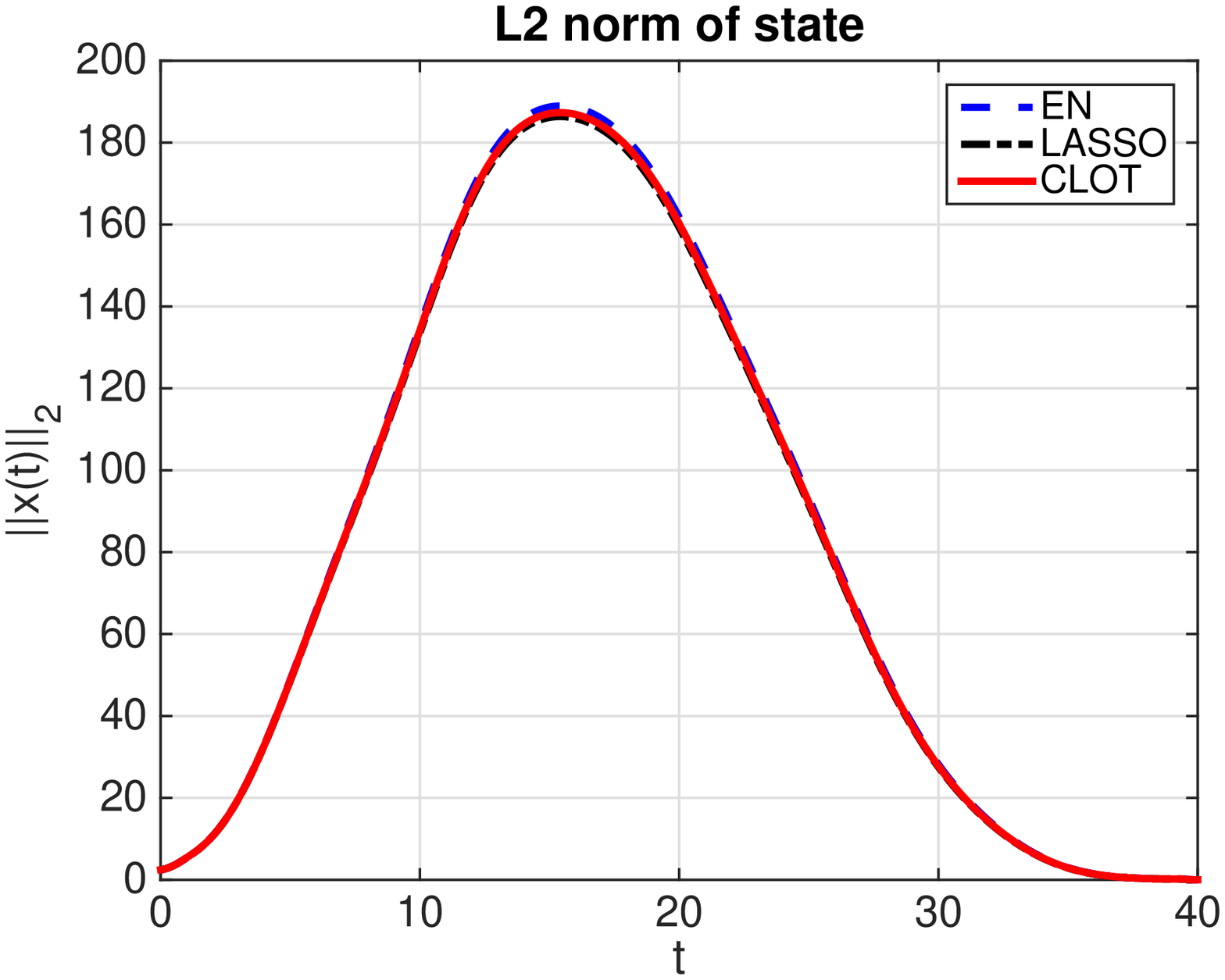}
\caption{State trajectory for the plant $P_6(s)$ with the initial
state $(1,1,1,1,1,1)^\top$ and $\l = 0.1$.}
\label{fig:P6_state}
\end{figure}

\begin{figure}[t]
\centering
\includegraphics[width=80mm]{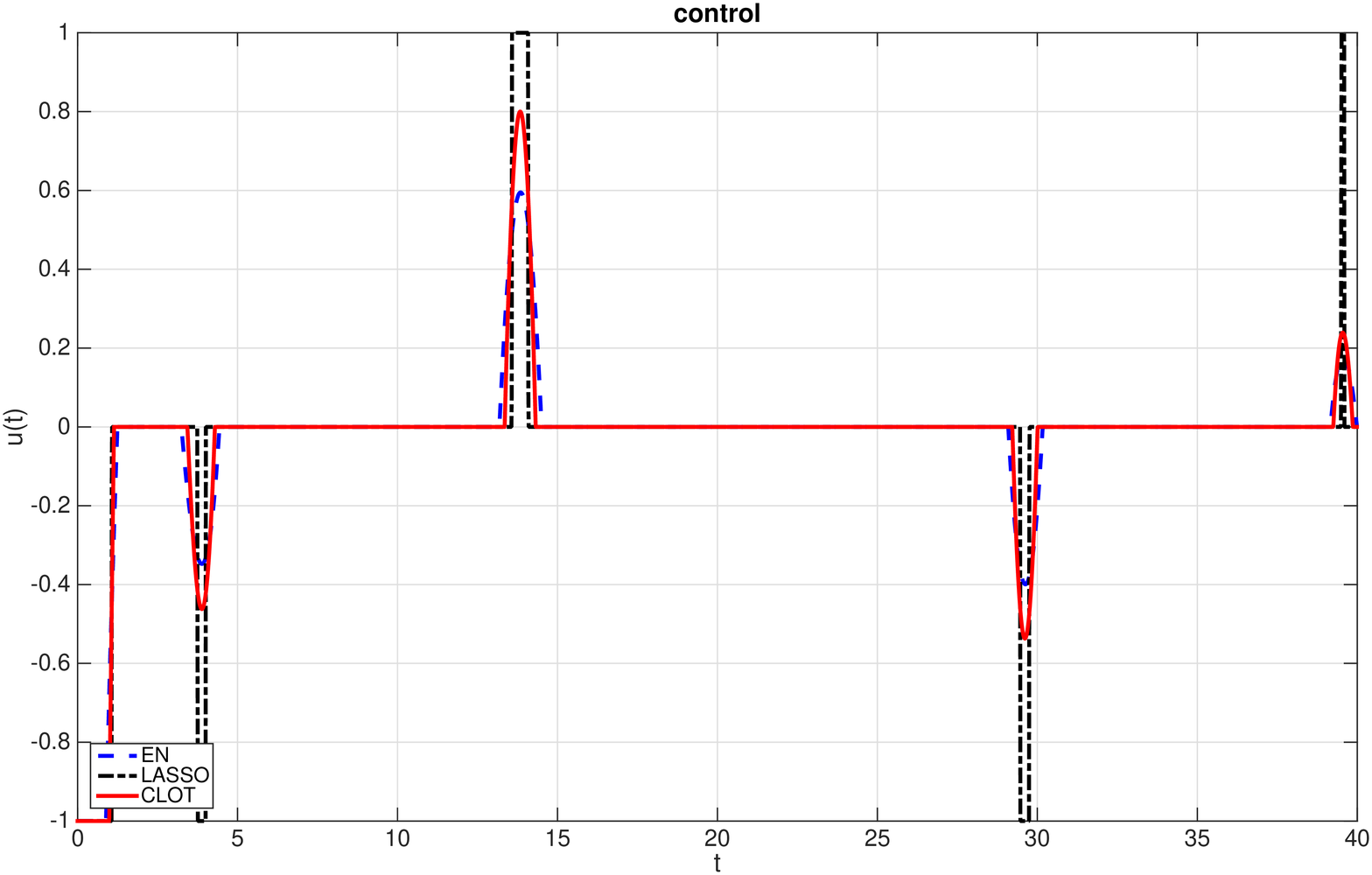}
\caption{Control trajectory for the plant $P_6(s)$ with the initial
state $(1,1,1,1,1,1)^\top$ and $\l = 0.1$.}
\label{fig:P6_control}
\end{figure}

\begin{figure}[t]
\centering
\includegraphics[width=80mm]{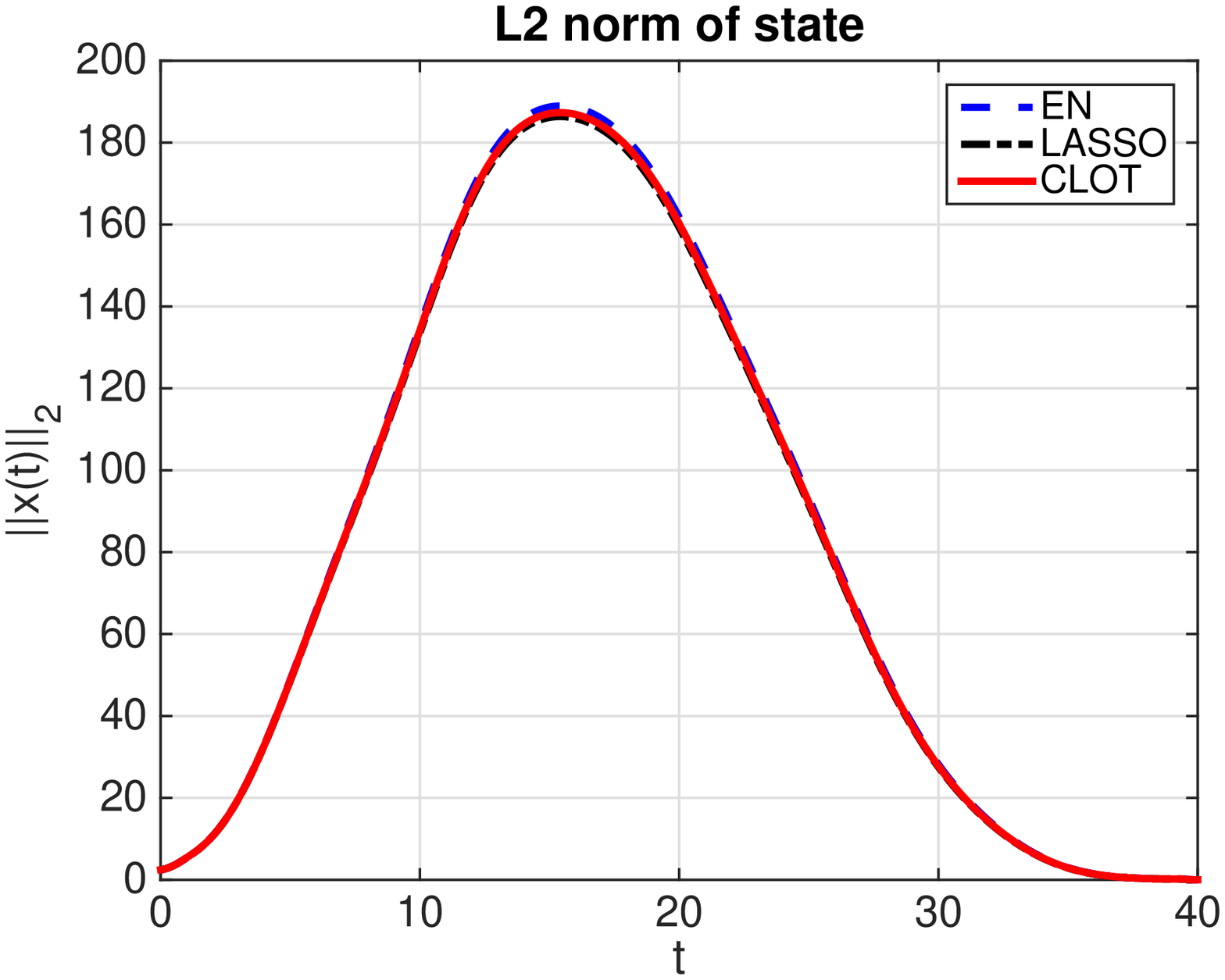}
\caption{State trajectory for the plant $P_7(s)$ with the initial
state $(1,1,1,1,1,1)^\top$ and $\l = 0.1$.}
\label{fig:P7_state}
\end{figure}

\begin{figure}[t]
\centering
\includegraphics[width=80mm]{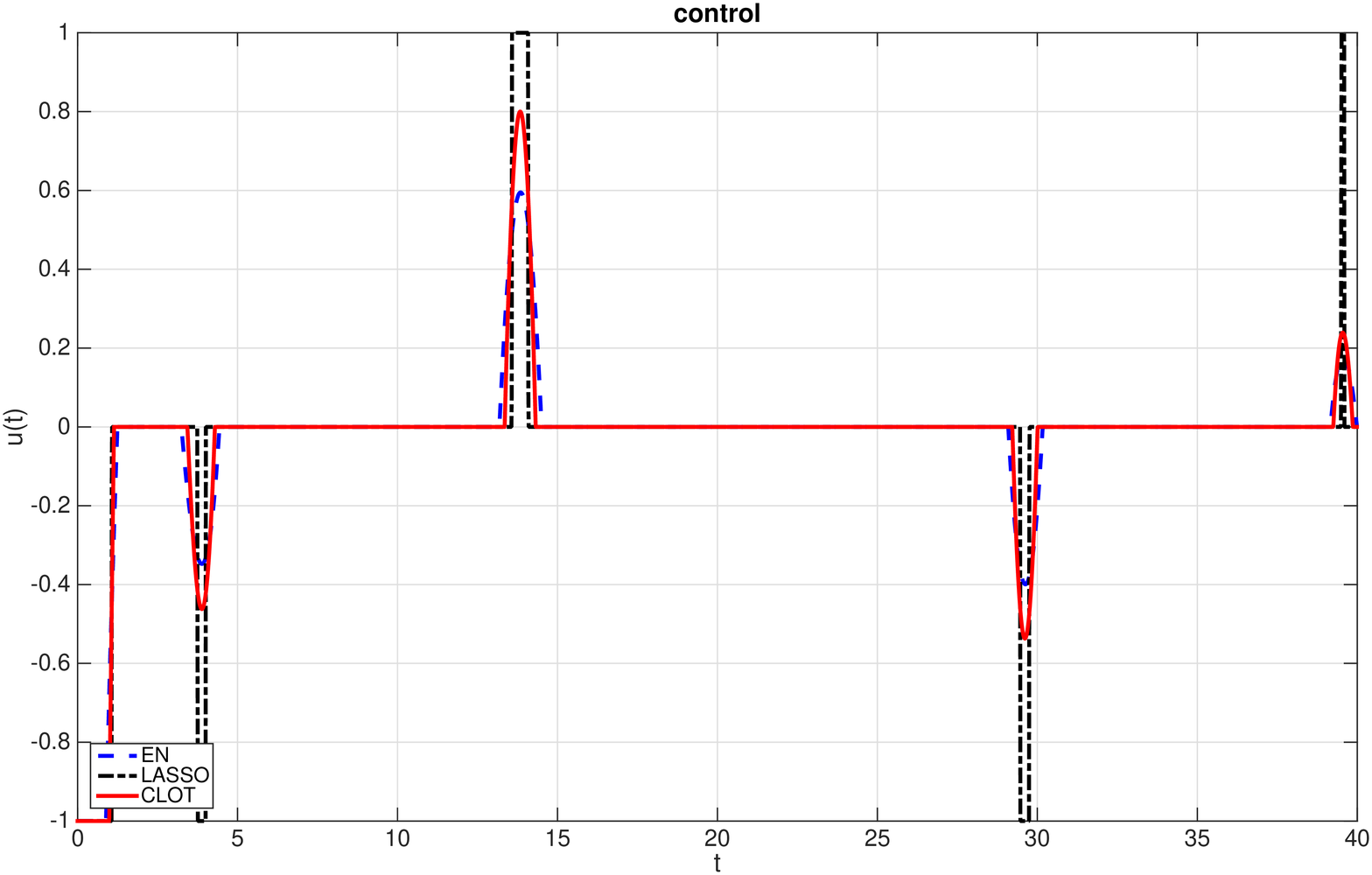}
\caption{Control trajectory for the plant $P_7(s)$ with the initial
state $(1,1,1,1,1,1)^\top$ and $\l = 0.1$.}
\label{fig:P7_control}
\end{figure}

To compare the sparsity densities of the three control signals,
we compute the fraction of time that each signal is nonzero.
In this connection, it should be noted that the LASSO control signal
is the solution of a \textit{linear programming} problem; consequently
its components \textit{exactly} equal zero at many time instants.
In contrast, the EN and CLOT control signals are the solutions of
\textit{convex} optimization problems.
Consequently, there are many time instants when the control signal
is ``small'' without being smaller than the machine zero.
Therefore, to compute the sparsity density, we applied a threshold of
$10^{-4}$, and treated a component of a control signal as being zero if
its magnitude is smaller than this threshold.
With this convention, the sparsity densities of the various control signals
are as shown in Table \ref{table:int}.
From this table it can be seen that the control signal generated using
CLOT norm minimization has significantly lower sparsity density
compared to that of EN, and is not much higher than that of LASSO.
Also, as expected, the sparsity density of LASSO does not change with $\l$,
whereas the sparsity densities of both EN and CLOT decrease as $\l$ is 
decreased.
For this reason, in other examples we present only the results for $\l = 0.1$.

\begin{table}
\centering
\btab{|l|r|r|r|}
\hline
\textbf{$\l$} & \textbf{LASSO} & \textbf{EN} & \textbf{CLOT} \\
\hline
$\l = 1$ & 0.1725 & 0.6050 & 0.5900 \\
        \hline
$\lambda = 0.1$ & 0.1725 & 0.3795 & 0.2665 \\ \hline
\etab
\caption{Sparsity indices of the control signals from various algorithms
for the plant $P_1(s)$ (fourth-order integrator) with the initial state
$(1,1,1,1)$.}
\label{table:int}
\end{table}
	
\subsection{Comparison of Sparsity Densities}

In this subsection we analyze the sparsity densities, that is, the
fraction of samples that are nonzero, using the three methods LASSO, EN,
and CLOT.
The advantage of using the sparsity \textit{density} instead of the sparsity
\textit{count} (the absolute number of nonzero entries) is that when
the sample time is reduced, the sparsity count would increase, whereas
we would expect the sparsity density to remain the same.
As explained above, we have applied a threshold of $10^{-4}$ in computing the
sparsity densities of various control signals.

Table \ref{table:spars} shows the sparsity densities for the nine
examples studied in Table \ref{table:plants}, in the same order.
From this table it can be seen that the CLOT norm-based control signal
is always more sparse than the EN-based control signal.
Indeed, in some cases the sparsity density of the CLOT control is
comparable to that of the LASSO control.
                                                   
\begin{table}		
\centering
\btab{|c|c|c|c|}
\hline
\textbf{No.} & \textbf{LASSO} & \textbf{EN} & \textbf{CLOT}\\
\hline
1 & 0.1690 & 0.5915 & 0.4475 \\
\hline
2 & 0.1690 & 0.3270 & 0.2480 \\ \hline
3 & 0.0480 & 0.1155 & 0.0830 \\ \hline
4 & 0.4055 & 0.5555 & 0.4225 \\ \hline
5 & 0.1655 & 0.3050 & 0.2180 \\ \hline
6 & 0.0040 & 0.0395 & 0.0805 \\ \hline
7 & 0.0595 & 0.1100 & 0.0845 \\ \hline
8 & 0.0568 & 0.1438 & 0.1125 \\ \hline
9 & 0.0568 & 0.1438 & 0.1125 \\ \hline
\etab
\caption{Sparsity densities for optimal controllers produced by
various methods}
\label{table:spars}
\end{table}

We also increased the number of samples from 2,000 to 4,000, and the
optimal values changed only in the third significant figure in almost
all examples for all three methods.
Therefore the figures in Table \ref{table:spars} are essentially equal to
the Lebesgue measure of the support set divided by $T$.

\section{Conclusions}
\label{sec:Conclusions}

In this article, we propose the CLOT norm-based control that minimizes the
weighted sum of $L^1$ and $L^2$ norms among feasible controls,
to obtain a continuous control signal that is sparser than the EN control
introduced in \citep{NagQueNes16}.
We have shown a discretization method, by which the CLOT optimal control
problem can be solved via finite-dimensional convex optimization.
Numerical experiments have shown the advantage of the CLOT control compared with
the LASSO and EN controls.
Future work includes the analysis of the sparsity and continuity of the CLOT control as a continuous-time signal.

\begin{ack}

The research of MN was supported in part by
JSPS KAKENHI Grant Numbers
15H02668, 
15K14006, and
16H01546.
The research of MV and NC was supported by the US National Science Foundation
under Award No. ECCS-1306630, the Cancer Prevention and Research
Institute of Texas (CPRIT) under award No.\ RP140517, and a grant from
the Department of Science and Technology, Government of India.

\end{ack}


\end{document}